\documentclass[aps,pre,floatfix,notitlepage,nofootinbib]{revtex4-1}

\usepackage{graphicx}
\usepackage{units,xspace}
\usepackage{color}
\usepackage{amsmath}

\newcommand{\dtheta}{\Delta\theta}
\newcommand{\eg}{{e.g., }}
\newcommand{\ie}{{i.e., }}

\newcommand{\Br}{{\bf r}}
\newcommand{\BR}{{\bf R}}

\newcommand{\BV}{{\bf V}}

\newcommand{\BF}{{\bf F}}

\newcommand{\CT}{\mathcal T}

\newcommand{\rad}{H}

\begin{document}

\title{Symmetry properties of nonlinear hydrodynamic interactions between responsive particles}
 
\author{Yulia Sokolov}

\author{Haim Diamant}

\affiliation{Raymond and Beverly Sackler School of Chemistry, and Center for Physics and Chemistry of Living Systems, Tel Aviv University, Tel Aviv 6997801, Israel}

\date{\today}

\begin{abstract}

Two identical particles driven by the same steady force through a viscous fluid may move relative to one another due to hydrodynamic interactions. The presence or absence of this relative translation has a profound effect on the dynamics of a driven suspension consisting of many particles. We consider a pair of particles which, to linear order in the force, do not interact hydrodynamically. If the system possesses an intrinsic property (such as the shape of the particles, their position with respect to a boundary, or the shape of the boundary) which is affected by the  external forcing, hydrodynamic interactions that depend nonlinearly on the force may emerge. We study the general properties of such nonlinear response. Analysis of the symmetries under particle exchange and under force reversal leads to general conclusions concerning the appearance of relative translation and the motion's time-reversibility. We demonstrate the applicability of the conclusions in three specific examples: (a) two spheres driven parallel to a wall; (b) two deformable objects driven parallel to their connecting line; and (c) two spheres driven along a curved path. The breaking of time-reversibility suggests a possible use of nonlinear hydrodynamic interactions to disperse or assemble particles by an alternating force.

\end{abstract}

\maketitle

\section{Introduction}
\label{sec_intro}

Colloidal suspensions are dispersions of nanometer-to-micron-sized objects
in a viscous fluid \cite{RusselBook}. Their collective dynamics is governed by
hydrodynamic interactions \cite{DhontBook}. These are velocity correlations
induced by the fluid flows that the motions of the objects
generate. Hydrodynamic interactions are the key factor in determining
such properties as the suspension's effective viscosity
\cite{RusselBook}, its density fluctuations \cite{DhontBook}, and velocity fluctuations \cite{Levine1998,Ramaswamy2001}.

In most practical cases the dynamics of the suspension is strongly
overdamped and the fluid flow can be assumed inertialess. This is
characterized by a negligible Reynolds number \cite{HappelBook,KimBook}. For example,
for micron-sized particles moving through water with a velocity of
order a micron per second, the Reynolds number is of order
$10^{-6}$. In this so-called Stokes limit, the response of the particles
to forces is linear and instantaneous. This instantaneous response depends
on the configuration of the particles while the configuration, in turn,
evolves due to the response. The resulting dynamics is nonlinear.
In addition, the hydrodynamic interactions are usually long-ranged 
and not pairwise-additive. As a result, even
for featureless, pointlike particles, the collective dynamics under an
external force can be extremely complicated \cite{Ramaswamy2001}.

In view of this complexity, it is helpful to identify cases where
hydrodynamic interactions are influential to larger or smaller
extents. One of the relevant questions is whether, for a pair of identical particles
driven by the same force, the instantaneous interaction affects only the velocity of
the pair's center of mass or also their relative velocity. The latter will
influence, for example, the rate of particle collisions \cite{Deutch} and
the stability of the collective dynamics \cite{Goldfriend2017,Witten2020},
affecting setups such as fluidized-bed chemical reactors.
Importantly, in the absence of relative velocity for pairs, the collective dynamics depends
nonlinearly on concentration \cite{Levine1998}, whereas in its presence the
dependence is linear \cite{Goldfriend2017}, \ie much stronger for dilute
suspensions.

Whether or not the relative translation velocity of a driven pair of particles
vanishes is a matter of symmetry. For an interaction that is
linear in the driving force, the relative velocity vanishes when the
pair's configuration is invariant to spatial inversion \cite{Goldfriend2016}. The
most common example is a pair of identical rigid spheres falling under
gravity in an unbounded viscous fluid. In the Stokes limit the spheres
affect each other's velocity and direction of motion but keep a
constant mutual distance \cite{HappelBook}. This follows from the fact that the
system remains unchanged if we invert the vector connecting the two
spheres. In the linear regime the
same conclusion can be reached based on the time-reversibility of
Stokes flow \cite{Purcell1977}. The instantaneous response makes all motions
quasistatic. In particular, reversing the direction of the force must
make the particles trace the same configurations backward. If they get
closer when the force is pointing down, they must get apart when the
force is pointing up. On the other hand, upon reversal of the force
the two spheres merely switch roles. Their relative velocity should
be the same. Therefore, it must be zero.

A well-known implication of Stokes' time-reversibility
relates to swimming at this overdamped limit \cite{Purcell1977,Elgeti2015}, which is very different from swimming at high Reynolds number. The problem is how to achieve net propulsion of an object out of cyclical changes in its configuration.
The forth and back parts of the cycle would cancel each other unless they constitute two nonreciprocal
motions. We pose here another problem related to time-reversibility: how to achieve net relative
motion of two or more objects under a cyclical alternating external drive. In the Stokes limit
a cycle of an alternating drive will not yield net relative motion. The particles will move back and forth, such that the configuration in the beginning of each cycle will be reproduced at its end.
If we wish to use an alternating force to move objects persistently together or apart, we
must depart from this limit.

The Stokes limit can be violated in various ways. Its quasistatic
property can be removed by inertia, viscoelasticity, or particle collisions \cite{Pine2005}.
A sufficiently strong driving force may set nonlinear effects due to the response of
either the fluid or the objects immersed in it. In the present work we focus on the latter. 
Rather than considering inert particles, we assume that they possess a certain
intrinsic property (\eg their shape, position within an external
potential or near a boundary, etc.), which changes under the driving force. This modifies
their individual dynamics as well as hydrodynamic interactions. A ubiquitous scenario involves deformable objects, such as vesicles, bubbles, emulsion droplets, and blood cells (see, \eg Ref.~\cite{Noguchi2005}). Considered as undeformed symmetric objects in
an unbounded fluid or along the axis of a tube, a pair of such driven objects would not develop relative velocity; yet, the deformed objects generally would. Since the
deformation is force-dependent, and the relative velocity of the
already deformed objects is proportional to the force, the effect is
necessarily nonlinear in the force. Such nonlinear effects play a role in 
several distinctive properties of blood flow \cite{KleinBook}.

From now on we will refer to the term ``hydrodynamic interaction'' to describe the
relative translation velocity of a pair of objects along the line that
connects them. Given a pair of identical objects under the same driving force, which {\it do not}
interact in the linear regime, we ask what we can say in general about their
interaction in the {\it nonlinear} regime based on symmetry alone. We
restrict the analysis to the nonlinear effect mentioned above. We neglect the transient
required for the objects to change their intrinsic property in
response to the steady force and consider the interaction between the already
modified objects in their steady state after the transient. We may compare this with the case of a nonpolar but polarizable particle interacting with a steady external field. Without polarizability there is no interaction. With polarizability, as soon as the field is turned on, the particle quickly attains a field-induced dipole moment, which then interacts with the field resulting in a nonlinear effect.  The phenomenon we describe is reminiscent of this case. This is in contrast to viscoelasticity, which has to do with the temporal {\it linear} response to a {\it time-dependent} force.

In Sec.~\ref{sec_gensym} we formulate the problem and infer the criteria for the
emergence of interaction and its dependence on the spatial configuration and force.
In Sec.~\ref{sec_examples} we give three illustrative examples (Sec.~\ref{sec_vis}) and three detailed examples (Secs.~\ref{sec_wall}, \ref{sec_spring} and \ref{sec_curve}) to demonstrate the applicability of the criteria. The system treated in Sec.~\ref{sec_wall} consists of two spherical particles held by a harmonic potential near a wall and driven parallel to the wall. In Sec.~\ref{sec_spring} we examine two deformable objects (each made of two spheres connected by a spring), driven along their connecting line. In the system of Sec.~\ref{sec_curve} two spherical particles are driven along a curved path (as in a circular optical vortex \cite{Sokolov2011}). In Sec.~\ref{sec_discussion} we summarize and discuss implications and extensions.

\section{General symmetry arguments}
\label{sec_gensym}

Let us consider two indistinguishable objects of arbitrary shape and properties. The vector $\BR$ connects the center of force of particle $2$ to the center of force of particle $1$. The particles are driven by an identical driving force $\BF$, which gives rise to steady-state velocities $\BV_1$ and $\BV_2$ of the two particles. The connection between the velocities and the force is given by response tensors ${\CT}_{ij}$,
\begin{subequations}
\begin{equation}
\BV_1(\Br,\BF)=\left[\CT_{11}(\BR,\BF) + \CT_{12}(\BR,\BF)\right]\cdot\BF,
\label{suchthat11}
\end{equation}
\begin{equation}
\BV_2(\Br,\BF)=\left[\CT_{21}(\BR,\BF) + \CT_{22}(\BR,\BF)\right]\cdot\BF,
\label{suchthat12}
\end{equation}
\label{suchthat01}
\end{subequations}
where each $\CT_{ij}$ is a $d \times d$ tensor in $d$ coordinates. These equations do not necessarily describe a linear response since any of these tensors may depend on $\BF$. Our purpose is to infer general properties of the velocity difference $\Delta \BV=\BV_1-\BV_2$ as a function of $\BR$ and $\BF$. Of particular interest is the relative velocity along the connecting line, $\Delta V_R \equiv \Delta \BV \cdot {\bf\hat{R}}$, which determines whether the particles attract ($\Delta V_R<0$), repel ($\Delta V_R>0$), or do not interact ($\Delta V_R=0$). The symmetries governing the linear response (where ${\CT}_{ij}$ is independent of $\BF$) are known (see, e.g.,  Refs.~\cite{Goldfriend2016, Witten2020}). We generalize this analysis to the nonlinear response (where ${\CT}_{ij}$ depends on $\BF$).

We begin with the exchange symmetry arising from the indistinguishability of the particles. The system is then invariant to the exchange $1 \leftrightarrow 2$ and $\BR \leftrightarrow -\BR$, such that
\begin{equation}
\CT_{11}(\BR,\BF)=\CT_{22}(-\BR,\BF);~\CT_{12}(\BR,\BF)=\CT_{21}(-\BR,\BF).
\label{suchthat02}
\end{equation}

We treat the symmetry of the system under inversion of $\BR$ by writing the tensors as a combination of $\BR$-even and $\BR$-odd contributions,
\begin{eqnarray}
&&\CT_{ij}=\CT_{ij}^e(\BR,\BF)+\CT_{ij}^o(\BR,\BF),\\
&&\CT_{ij}^e(-\BR,\BF)=\CT_{ij}^e(\BR,\BF);~\CT_{ij}^o(-\BR,\BF)=-\CT_{ij}^o(\BR,\BF).\nonumber
\label{suchthat03}
\end{eqnarray}
Following Eq.~(\ref{suchthat01}) we get for the $\BR$-even contributions
\begin{equation}
\BV_1^e(\BR,\BF)=\BV_1^e(-\BR,\BF)=\BV_2^e(\BR,\BF),
\label{suchthat04}
\end{equation}
and for the $\BR$-odd contributions, 
\begin{equation}
\BV_1^o(\BR,\BF)=-\BV_1^o(-\BR,\BF)=-\BV_2^o(\BR,\BF).
\label{suchthat05}
\end{equation}
Thus $\BR$-even contributions do not lead to relative velocity ($\Delta \BV^e\cdot{\bf\hat{R}}=0$). Any nonzero relative velocity must arise from the $\BR$-odd contributions ($\Delta V_R=\Delta \BV^o\cdot{\bf\hat{R}}$). Note that unlike the argument given in Sec.~\ref{sec_intro} based on time-reversibility, this argument does not require linear response.

We now discuss the symmetry under the inversion of $\BF$. We split each of the $\BR$-even and $\BR$-odd contributions further into $\BF$-even and $\BF$-odd terms. There are four cases, $\CT^{*\#}$ ($*$ stands for $e$ or $o$ for $\BR$ and $\#$ stands for $e$ or $o$ for $\BF$), for each of the four response tensors $\CT_{ij}$ ($i,j=1,2$) --- overall sixteen functions of $\BR$ and $\BF$:
\begin{eqnarray}
\label{suchthat06}
&&\CT_{ij}=\sum_{*,\#}\CT_{ij}^{*\#}(\BR,\BF),\\
&&\CT^{*e}_{ij}(\BR,-\BF)=\CT^{*e}_{ij}(\BR,\BF);~\CT^{*o}_{ij}(\BR,-\BF)=-\CT^{*o}_{ij}(\BR,\BF).\nonumber
\end{eqnarray}
The contributions $\CT^{e\#}_{ij}$ do not cause relative velocity, as concluded above. Therefore we focus on the contributions $\CT^{o\#}$. For $\CT^{oo}_{ij}$,
\begin{subequations}
\begin{equation}
\CT^{oo}_{11}(\BR,-\BF)=-\CT^{oo}_{11}(\BR,\BF)=\CT^{oo}_{11}(-\BR,\BF)=\CT^{oo}_{22}(\BR,\BF),\\
\label{suchthat71}
\end{equation}
\begin{equation}
\CT^{oo}_{12}(\BR,-\BF)=-\CT^{oo}_{12}(\BR,\BF)=\CT^{oo}_{12}(-\BR,\BF)=\CT^{oo}_{21}(\BR,\BF).\label{suchthat72}
\end{equation}
\label{suchthat07}
\end{subequations}
This implies [see Eq.~(\ref{suchthat01})] that the $oo$ tensors lead to
\begin{eqnarray}
&&\BV_1(\BR,-\BF)=-\BV_2(\BR,\BF),\nonumber\\
&&\Delta \BV(-\BF)=\Delta \BV(\BF),
\label{suchthat08}
\end{eqnarray}
i.e., to an interaction that is even in $\BF$. The physical implication is that, upon the reversal of the force, attraction remains attraction ($\Delta V_R<0$) and repulsion remains repulsion ($\Delta V_R>0$). In such a case the nonlinear response breaks the time-reversal symmetry of the Stokes flow. 

The last case is $\CT^{oe}_{ij}$, where, in a similar way, we find
\begin{eqnarray}
&&\BV_1(\BR,-\BF)=\BV_2(\BR,\BF),\nonumber\\
&&\Delta \BV(-\BF)=-\Delta \BV(\BF),
\label{suchthat09}
\end{eqnarray}
meaning that the interaction is odd in $\BF$. In this case the reversal of the force will turn an attraction into repulsion and vice versa, without breaking time-reversibility.

In the special case of linear response, where $\CT$ is independent of $\BF$, $\CT^{*o}$ all vanish and any interaction must arise from $\CT^{oe}$, implying time-reversibility.

Equation (\ref{suchthat06}) describes the common general case, where the response contains all terms. By decomposing a general response into symmetry-based contributions as defined above, we can identify the effects that are responsible for hydrodynamic interactions. This will be illustrated in the following sections. The decomposition into $\CT^{*\#}$ is useful, for instance, in the multipole expansion of the hydrodynamic interaction between two well-separated objects. In such an expansion each consecutive term involves another gradient of the interaction kernel with respect to $\BR$. As a result one gets alternating $\BR$-even and $\BR$-odd terms, the latter leading to hydrodynamic interactions, as already noted for linear interactions \cite{Goldfriend2016}. For nonlinear interactions each of these multipoles can be further split into $\BF$-odd and $\BF$-even terms, which preserve and break time-reversibility, respectively.

We now present three visual and three detailed examples.

\section{Examples}
\label{sec_examples}

Focusing on hydrodynamic interactions arising due to nonlinear response, we treat systems that to linear order in $\BF$ have no interaction ($\Delta V_R=0$). As we have just shown, this implies that $\CT_{ij}(\BR,\BF=0)=\CT_{ij}(-\BR,\BF=0)$.

All the examples presented in this section exhibit a nonlinear effect that breaks the inversion symmetry in some way, leading to hydrodynamic interactions. We begin with a few illustrative examples to demonstrate the application of the symmetry arguments in cases where detailed analysis is very difficult. We then proceed to explicit treatments of three examples, with increasing order of accuracy.

\subsection{Illustrative examples}
\label{sec_vis}

\begin{figure}[tbh]
  \centering
  \includegraphics[width=0.5\columnwidth]{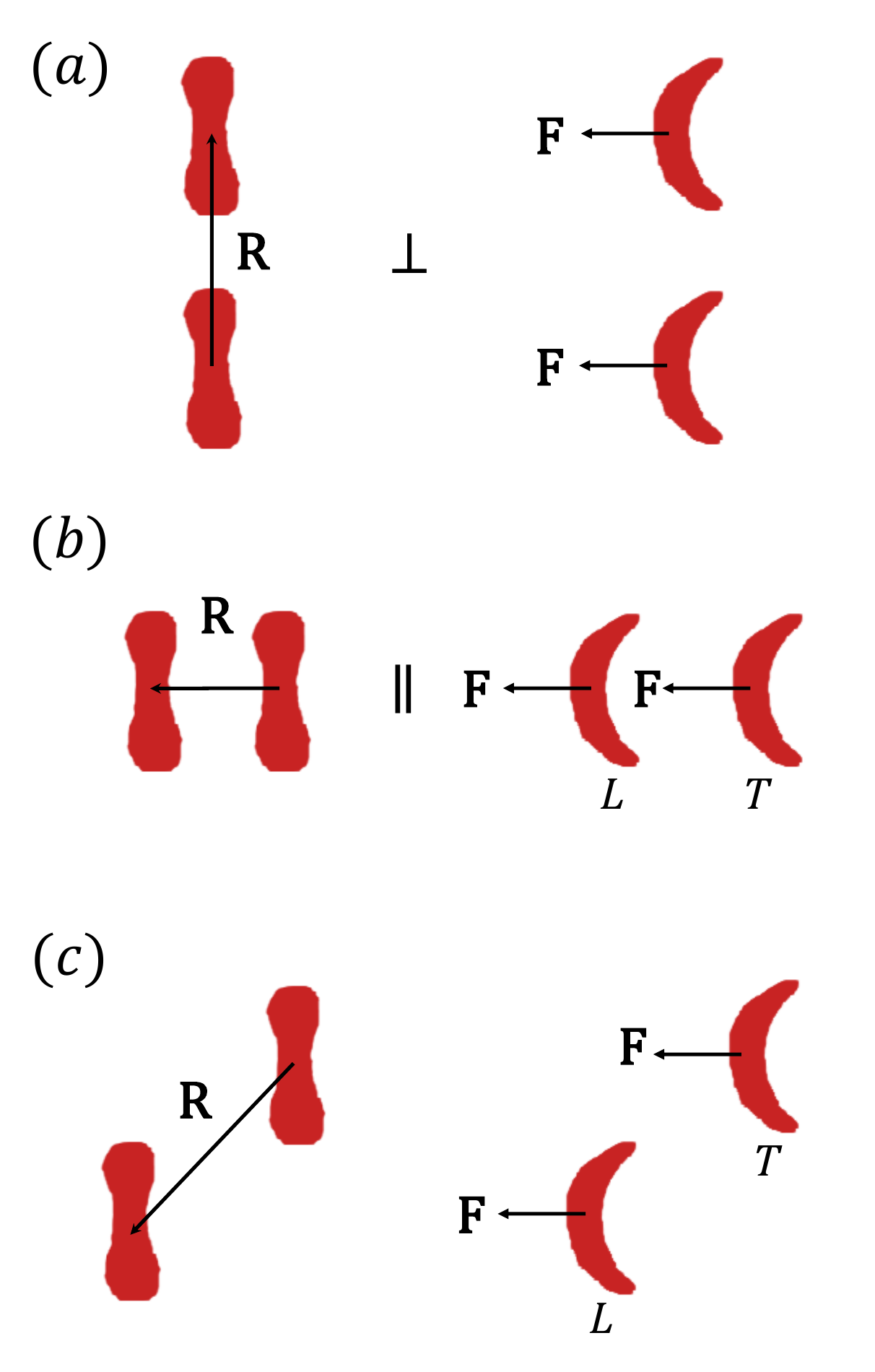}
\caption{First example of nonlinear hydrodynamic interaction: two flexible objects deforming under an external force. In the absence of force the shapes of the objects are symmetric and they do not interact hydrodynamically (left-hand images). (a) The force is applied perpendicular to the vector connecting the centers of force of the objects. In this configuration the deformed objects do not interact. (b) The force is applied parallel to the connecting vector. Here the objects interact and the interaction is even in $\BF$, breaking time-reversal symmetry. (c) For a general relative orientation between $\BF$ and $\BR$ the qualitative conclusions for (b) still hold.}
\label{fig:ill1}
\end{figure}

In Fig.~\ref{fig:ill1} we show two flexible objects (representing, for example, red blood cells), which in the absence of force are invariant to $\BR$-inversion. This means that they do not interact to linear order. In the presence of the force $\BF$ the objects deform identically and lose their fore-aft symmetry (we neglect any correlated deformations). In the case of perpendicular alignment, $\BF \perp \BR$ [Fig.~\ref{fig:ill1}(a)] the configuration is still $\BR$-even, and we can immediately conclude that nonlinear interaction does not appear. In the case of parallel alignment, $\BF \parallel \BR$ [Fig.~\ref{fig:ill1}(b)], the $\BR$-inversion symmetry is broken, with a difference between the leading ($L$) and trailing ($T$) objects. Therefore, following Eq.~(\ref{suchthat05}), they should move relatively. Without a detailed analysis one cannot know whether these objects attract or repel. In Sec.~\ref{sec_spring} we treat a simple example of such a configuration of deformable objects.

What we can predict is the behavior of the system under force inversion and therefore its time-reversal symmetry. When the direction of the force is inverted, the deformations reverse as well. The leading particle becomes trailing and vice versa, and the relative velocity remains the same. This scenario fits the description in Eqs.~(\ref{suchthat07}) and (\ref{suchthat08}). The velocity difference is even in $\BF$ ($\CT_{ij}$ is odd in $\BF$), meaning that if the objects repel/attract, they will also repel/attract once the force is inverted. This breaks time-reversibility. In a case of a general angle between $\BR$ and $\BF$ and of identical orientation, as shown in Fig.~\ref{fig:ill1}(c), the $\BR$-inversion symmetry is still broken, while the symmetry with respect to $\BF$-inversion still holds. Hence, all the conclusions reached for the parallel alignment remain valid. In examples \ref{fig:ill1}(a) and \ref{fig:ill1}(c) there is  rotational interaction which in the next instant will lead to different orientations of the two objects. Yet, as we have emphasized in the Introduction, the very existence or absence of the {\it instantaneous} translational interaction of the idealized configuration has a profound impact on the collective dynamics of many such objects.

\begin{figure}[tbh]
  \centering
  \includegraphics[width=0.5\columnwidth]{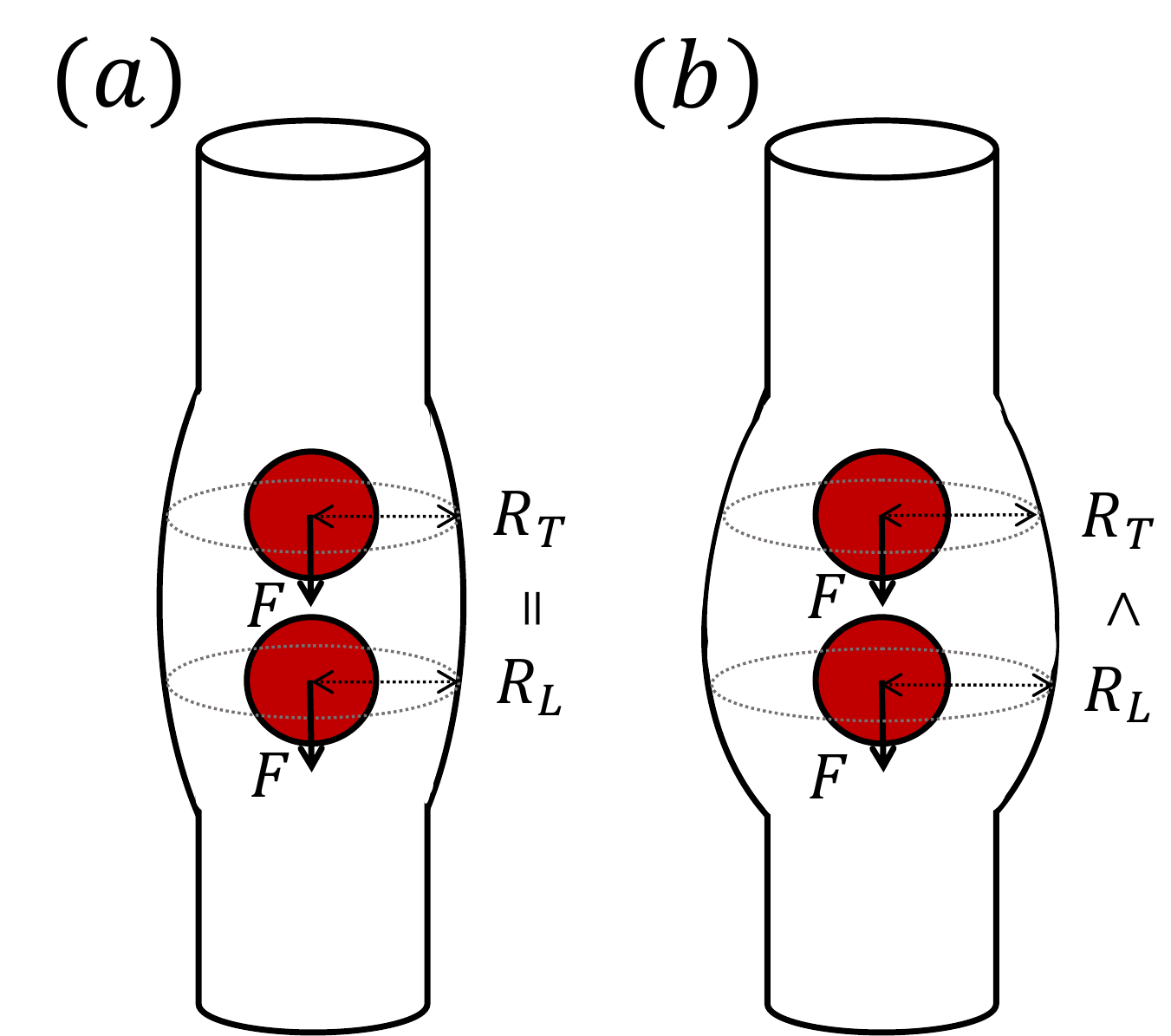}
\caption{Second example of nonlinear hydrodynamic interaction: two rigid symmetric objects driven along a deformable tube. (a) The particles cause a symmetric tube deformation, leading to no interaction. (b) The leading particle causes a larger increase in tube diameter, thus moving faster than the trailing one (repulsion). The interaction is $\BF$-even and therefore time-irreversible.}
\label{fig:ill2}
\end{figure}

Another visual example is shown in Fig.~\ref{fig:ill2}. Here two rigid and symmetrical objects move along the central axis of a tube (for instance, two drug-carrying particles in a narrow vein). Similar to the previous example, there is no linear interaction between the objects. In the case of a symmetric deformation of the tube, Fig.~\ref{fig:ill2}(a), the system remains $\BR$-even and therefore, there is no interaction to all orders in $\BF$. In the case of asymmetric deformation, where the diameter of the tube around a leading particle ($L$) is different from (say, larger than) that around the trailing particle ($T$) [Fig.~\ref{fig:ill2}(b)], the $\BR$-inversion symmetry is broken. For example, they will repel due to the differences in the friction with the boundary. The system is $\BF$-even and, therefore, time-irreversible.

In the asymmetric cases of both examples, the tensors $\CT_{ij}$ are purely odd in $\BF$ ($\Delta V_R$ even in $\BF$). This means that any detailed calculation of the tensors for these complex problems will inevitably produce odd powers of $\BF$ only.

\begin{figure}[tbh]
  \centering
  \includegraphics[width=0.5\columnwidth]{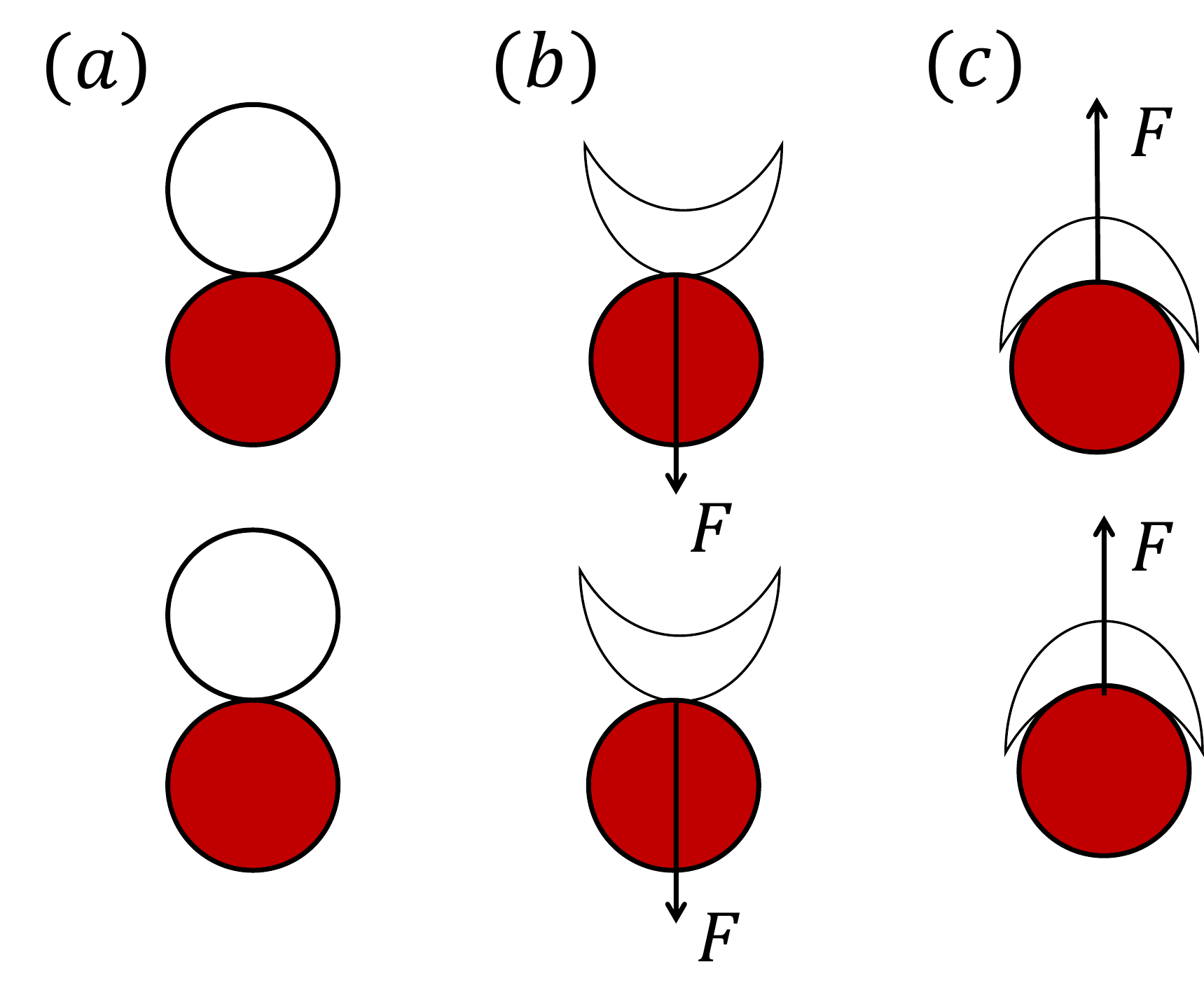}
\caption{Third example of nonlinear hydrodynamic interaction: two dimers made each of a rigid particle (lower, red) and a deformable one (upper, white). (a) When no force is applied the objects are symmetric. (b) When force is applied the white part of a dimer deforms and the dimer becomes asymmetric, leading to nonlinear hydrodynamic interaction. (c) When the force is reversed, the white part of the dimer deforms in the opposite direction, creating a dimer of different asymmetric shape. Here too the two objects will interact, but differently, and the interaction has no definite parity with respect to $\BF$.}
\label{fig:ill3}
\end{figure}

In general cases, $\CT_{ij}$ does not have a definite parity with respect to $\BF$, and the change in the relative velocity upon $\BF$ reversal will not be easily predicted. Figure \ref{fig:ill3} shows such an example, where, upon force reversal, the objects not only switch roles as leading and trailing, but also change their shapes according to the force direction.

\subsection{Detailed example 1: two spheres driven parallel to a wall}
\label{sec_wall}

\subsubsection{Model}
\label{sec_modelwall} 

\begin{figure}[tbh]
  \centering
  \includegraphics[width=0.5\columnwidth]{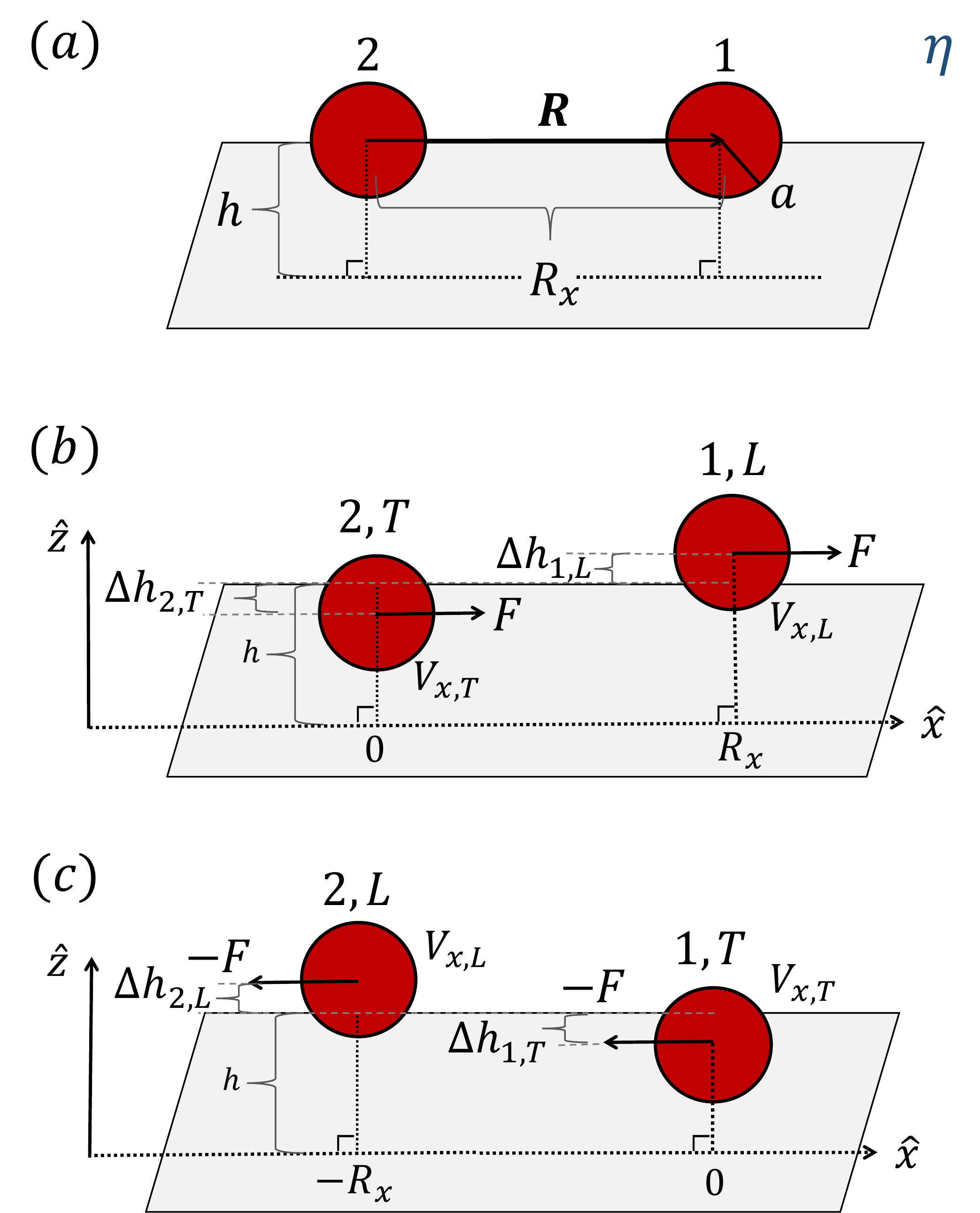}
\caption{Two spherical particles driven parallel to a wall. (a) In the absence of force the particles are held by an external potential at a distance $h$ from the wall. (b) Under the force the particles move in the ${\bf\hat{x}}$ direction and are displaced in the ${\bf\hat{z}}$ direction. The leading particle ($L$) is displaced away from the wall, and the trailing one ($T$) approaches the wall. A restoring harmonic force perpendicular to the wall limits these displacements. (c) Upon the reversal of the force and reaching the new steady state the particles switch roles.}
\label{fig:modelwall}
\end{figure}

In this section we consider two rigid spheres $1$ and $2$ of radius $a$, which in the absence of external force are held at the same distance $z=h$ from a rigid wall [Fig.~\ref{fig:modelwall}(a)]. The system is immersed in a fluid of viscosity $\eta$. We take the ${\bf\hat{x}}$ axis along the line connecting the two spheres and the ${\bf\hat{z}}$ axis perpendicular to the wall. The two particles are separated by the vector $\BR=\BR_1-\BR_2=R_x{\bf\hat{x}}$. The particles are steadily driven by a force $\BF=F{\bf\hat{x}}$ [see Fig.~\ref{fig:modelwall}(b)]. The direction of the force defines the leading ($L$) and the trailing ($T$) particles. The two particles move in the ${\bf\hat{x}}$ direction with velocities $V_{x,1}$ and $V_{x,2}$. We determine how the particles interact when driven parallel to the wall.

To begin with, we note that, when restricted to the two-dimensional plane, the motion does not break $\BR$-inversion symmetry and therefore does not create relative velocity parallel to the wall, i.e., $V_{x,1}=V_{x,2}$. Thus, to linear order in $\BF$ there is no interaction parallel to the wall.

What breaks the symmetry of this system is the fact that due to the presence of the wall, the parallel driving also leads to perpendicular motion \cite{HappelBook}. This motion in the ${\bf \hat{z}}$ direction is resisted by an external restoring force $F_z$, such that at steady state the particles are displaced from their initial distance $h$ from the wall to $h_i=h+\Delta h_i$, $i=1,2$ [see Fig.~\ref{fig:modelwall}(b)]. For simplicity we assume a spring-like restoring force, $F_{z,i}=-k \Delta h_i$. The configuration of the tilted particles is no longer $\BR$-even. This might lead to relative velocity parallel to the wall, $\Delta V_R=V_{x,1}-V_{x,2}\neq 0$.

The specific mechanism that we consider relies on the dependence of the particles' self-mobilities on their distance from the wall. Since they are tilted, their self-mobilities differ and so do their velocities parallel to the wall.

Regardless of the specific mechanism which causes the relative velocity, based on symmetry, if the force is reversed in direction, the two particles switch their roles as leading and trailing [See Fig.~\ref{fig:modelwall}(c)]. At the resulting steady state, the tilt will be reversed, $\Delta h_1(-\BF)=\Delta h_2(\BF)$, the velocities will switch sign and exchange, $V_{x,2}(-\BF)=-V_{x,1}(\BF)$, and the relative velocity $\Delta V_R$ will stay the same. Thus the interaction is $\BF$-even.

We now treat the problem in detail.

\subsubsection{Results}
\label{sec_reswall}

The system contains three intrinsic length scales, $a$, $h$ and $F/k$. The following approximate calculation is performed to the leading order in small $a/h$ and $F/(kh)$. The latter limit, of strong confining force, implies arbitrarily small $\Delta h_i$. Our calculation allows any value of $h/R_x$. In addition we assume $a \ll R_x$. This allows us to replace the required components of the response tensors as follows (the Stokeslet approximation):
\begin{eqnarray}
&&(\CT_{11})_{xx} \simeq B_{x,1}(h_1),\ \ (\CT_{22})_{xx} \simeq B_{x,2}(h_2),\nonumber \\
&&(\CT_{12})_{xx} \simeq G_{xx}(R_x),\ \ (\CT_{21})_{xx} \simeq G_{xx}(-R_x),\nonumber \\
&&(\CT_{12})_{zx} \simeq G_{zx}(R_x),\ \ (\CT_{21})_{zx} \simeq G_{zx}(-R_x).
\label{walltensors}
\end{eqnarray}
Here
\begin{equation}
B_x(z) =\frac{1-(9/16)(a/z)}{6\pi\eta a}
\label{Bsx}
\end{equation}
is the self-mobility parallel to the wall, which depends on the distance from the wall $z$ \cite{HappelBook}, neglecting terms of order $(a/z)^3$ and higher. To leading order in $\Delta h_i$ we assume first that the hydrodynamic interaction is between particles positioned at the same distance $h$ from the wall. The hydrodynamic interaction parallel to the wall is then \cite{PozrikidisBook}
\begin{equation}
G_{xx}(R_x)=G_{xx}(-R_x)=\frac{1}{4 \pi \eta} \left(\frac{1}{|R_x|}-\frac{R_x^2}{\rho^3}-\frac{12 h^4}{\rho^5}\right),
\label{Gxx}
\end{equation}
where $\rho=\sqrt{R_x^2+4h^2}$ is the distance between one particle and the ``image'' of the other behind the wall. The interaction causing the perpendicular velocity is \cite{PozrikidisBook}
\begin{equation}
G_{zx}(R_x)=-G_{zx}(-R_x)=\frac{3}{2 \pi \eta} \frac{h^3 R_x}{\rho^5}.
\label{Gzx}
\end{equation}
The velocity of each particle parallel to the wall, affected by the forces $F$ on itself and on the other particle, is then
\begin{eqnarray}
V_{x,1}&=&\left[B_{x,1} + G_{xx}(R_x)\right]F,\nonumber\\
V_{x,2}&=&\left[B_{x,2} + G_{xx}(-R_x)\right]F.
\label{Vx}
\end{eqnarray}

We see in Eq.~(\ref{Gxx}) that $G_{xx}$ is even in $R_x$, and therefore, according to Eq.~(\ref{Vx}), there is no interaction parallel to the wall unless $B_{x,1} \neq B_{x,2}$ \footnote{In fact, for this system $G_{xx}$ is symmetric under the inversion of particle positions also when $h_1 \neq h_2$ (see Appendix A).}. Since at different distances from the wall the self-mobilities are not the same [see Eq.~(\ref{Bsx})], the particles do interact.

We now turn to the motion perpendicular to the wall. If the particles were free to move in the ${\bf \hat{z}}$ direction, their perpendicular velocities would be
\begin{eqnarray}
V_{z,1}&=&G_{zx}(R_x)F,\nonumber\\
V_{z,2}&=&G_{zx}(-R_x)F.
\label{Vz}
\end{eqnarray}
Since $G_{zx}$ is $R_x$-odd [Eq.~(\ref{Gzx})], these velocities are equal and opposite.  The existence of the restoring force $F_z$ makes the perpendicular movement finite. The leading particle is displaced by $\Delta h_L >0$ away from the wall, and the trailing one by $\Delta h_T<0$ toward the wall. This is the symmetry breaking that causes the nonlinear interaction along the ${\bf\hat{x}}$ direction such that $\Delta V_R \neq 0$. The hydrodynamic forces that oppose the restoring forces $F_{z,i}=-k \Delta h_i$ are given by $V_{z,i}/B_z(h)$ [neglecting corrections of order $(a/z)^2$ and higher], where 
\begin{equation}
B_z(z)=\frac{1-\frac{9}{8}(a/z)}{6 \pi \eta a}
\label{Bsz}
\end{equation}
is the mobility perpendicular to the wall \cite{HappelBook}, to linear order in $a/z$. This balance gives
\begin{equation}
\Delta h_L=\frac{9 a h^3 F}{k} \frac{R_x}{\rho^5},\  \Delta h_T=-\Delta h_L.
\label{h}
\end{equation}

\begin{figure}[tbh]
  \centering
  \includegraphics[width=0.5\columnwidth]{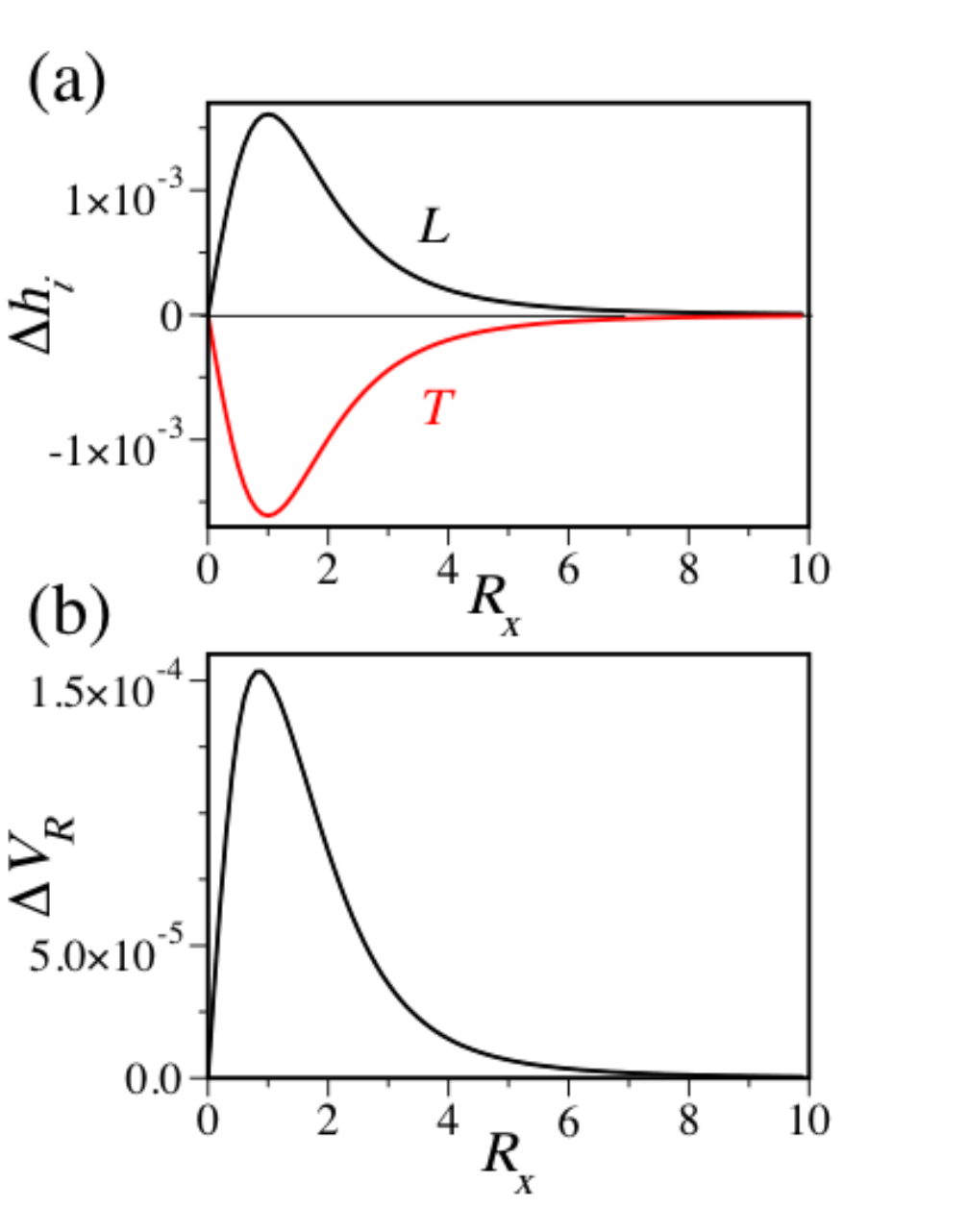}
\caption{(a) The perpendicular displacements of the leading ($L$) and trailing ($T$) particles as a function of their mutual distance. (b) The hydrodynamic interaction (relative velocity) as a function of distance. The results are given in units where $h=F=\eta\equiv 1$. In addition, $a=0.1$ and $k=10$.}
\label{fig:reswall}
\end{figure}

Figure \ref{fig:reswall}(a) shows the two perpendicular displacements as a function of the particle separation. The displacements are antisymmetric. This is because they are the consequence of the $R_x$-odd hydrodynamic interaction $G_{zx}$. The displacements exhibit a nonmonotonous dependence on particle separation, with a maximum at $R_x=h$. This arises from the interplay between the effect of the wall and the strength of interaction. When $R_x \ll h$ the symmetry-breaking effect of the wall becomes negligible. When $R_x \gg h$, the hydrodynamic interaction weakens. The strength of the restoring force, determined by $k$, affects the amplitude of the displacements $\Delta h_i$.

We substitute $z=h_i=h+\Delta h_i$ from Eqs.~(\ref{h}) in Eq.~(\ref{Bsx}) to obtain the parallel self-mobilities, $B_{x,i}$. Since $G_{xx}$ does not contribute to the relative velocity, we simply have $V_{x,L}-V_{x,T}=(B_{x,L}-B_{x,T})F$. The interaction in the ${\bf{\hat x}}$ direction is then
\begin{equation}
\Delta V_R=\frac{27 a h F^2}{16\pi\eta k} \frac{(\rho^7+32 h^3 R_x^4+128 h^7) |R_x|}{\rho^{12}}.
\label{dVR}
\end{equation}
The interaction is repulsive as anticipated. Unlike the usual case where the hydrodynamic interaction decays monotonously with distance, in this case the interaction is nonmonotonous in $R_x$, as can be seen in Fig.~\ref{fig:reswall}(b). This follows from the nonmonotonous behavior of the displacements $\Delta h_i$. Recall that $\Delta V_R$ vanishes in an unbounded fluid and therefore must vanish in the limit $R_x/h \rightarrow 0$. On the other hand, when $R_x$ becomes larger than $h$ we can observe the usual decay of the interaction with distance, in the present case as $1/R_x^4$. Curiously, the repulsion for $R_x<h$ {\it strengthens} with increasing distance. This implies that particles in this range will accelerate away from each other.

Equation (\ref{dVR}) demonstrates the features described in Sec.~\ref{sec_gensym}. The force modifies the intrinsic property of the particles, namely, their distance from the wall. This breaks the $\BR$-inversion symmetry, creating nonlinear interaction parallel to the wall. The symmetry of the problem under force reversal leads to interaction which is even in $F$. Here the result is proportional to $F^2$ due to the leading-order approximation that we have employed. The arguments in Sec.~\ref{sec_gensym}, Eq.~(\ref{suchthat08}), ensure that a more accurate calculation  will necessarily give higher but only even powers in $F$. Such higher orders will be demonstrated in the following examples.

\subsection{Detailed example 2: two spring-like objects}
\label{sec_spring}

\subsubsection{Model}
\label{sec_modelspring}

\begin{figure}[tbh]
  \centering
  \includegraphics[width=0.5\columnwidth]{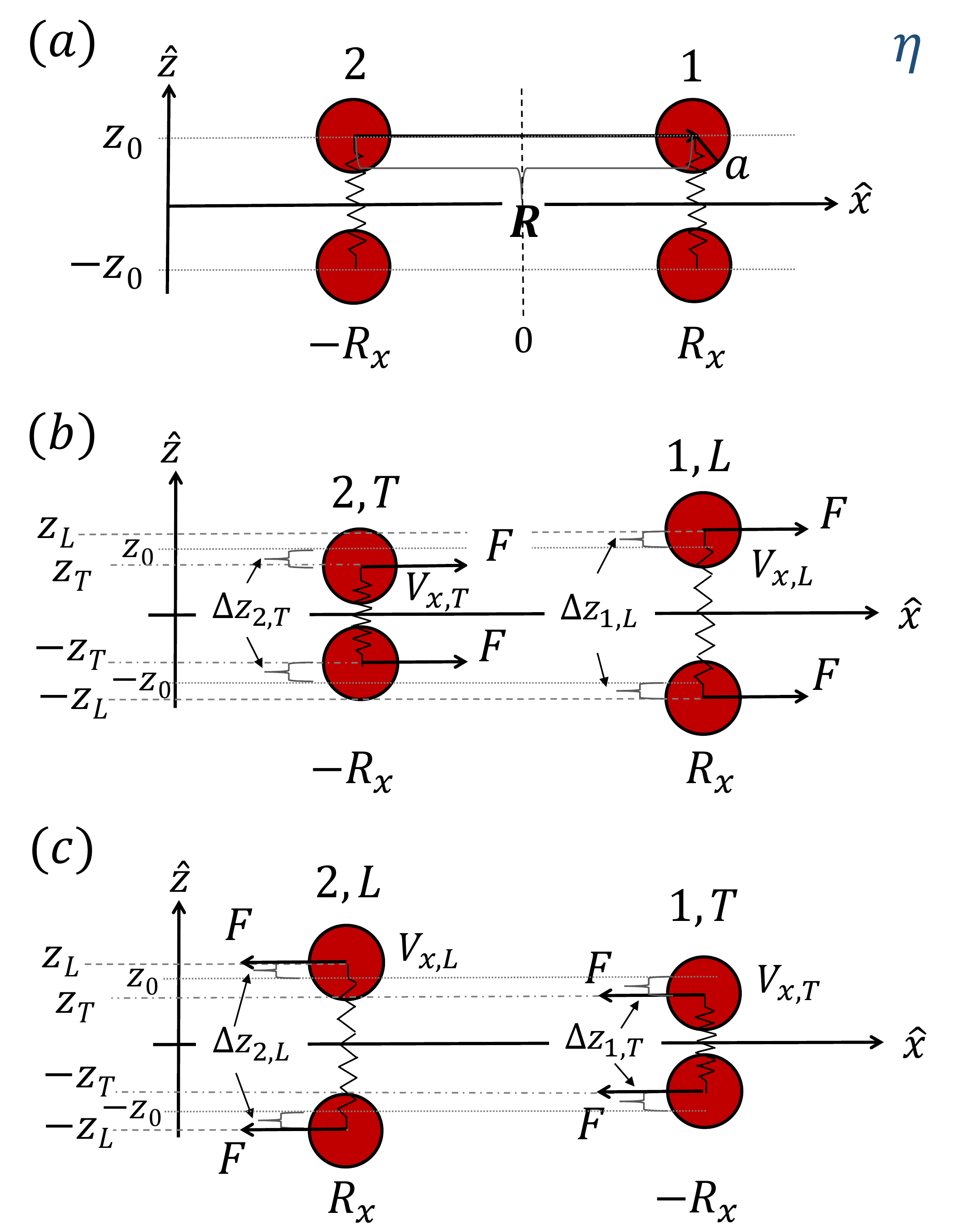}
\caption{Two driven deformable objects. Each object is made of two spheres connected by a spring. (a) In the absence of force the springs are relaxed and the objects are identical. (b) Under the force the centers of the objects move in the force direction (${\bf\hat{x}}$), while the spheres also move in the perpendicular direction (${\bf\hat{z}}$) -- the leading ($L$) object elongates and the trailing ($T$) shrinks. (c) Upon reversal of the force and reaching the new steady state, the particles switch roles.}
\label{fig:springmodel}
\end{figure}

To demonstrate the effect of the objects' deformability, mimicking the scenario of Fig.~\ref{fig:ill1}(b), we consider the two idealized spring-like objects $1$ and $2$ shown in Fig.~\ref{fig:springmodel}, immersed in a viscous fluid of viscosity $\eta$. Each object is made of two small spheres of radius $a$ connected by an infinitely thin spring of spring constant $k$ and equilibrium length $2z_0$. The spheres within each object $i$ are aligned in the ${\bf\hat{z}}$ direction and positioned at $\pm z_i$. The centers of the two objects are separated along the ${\bf\hat{x}}$ direction by the vector $\BR=\BR_1-\BR_2=R_x{\bf\hat{x}}$. An identical force $\BF=F{\bf\hat{x}}$ is steadily applied to all four spheres [See Fig.~\ref{fig:springmodel}(b)]. In response to the force the two particles develop velocities in the ${\bf\hat{x}}$ direction, $V_{x,1}$ and $V_{x,2}$. Our aim is to determine whether the particles interact (attract or repel) and discuss the symmetry underlying this interaction.

If the springs were infinitely rigid the system would be invariant to $\BR$-inversion and would not develop relative motion, $V_{x,1}=V_{x,2}$. This implies that, to linear order in $\BF$, the two objects do not interact.

The extra degrees of freedom in the ${\bf\hat{z}}$ direction due to the springs allow the spheres within each object to get closer or farther apart. Perceiving the flow lines produced by the driven spheres, we see that the leading spheres move the trailing pair together, while the trailing spheres move the leading ones apart. All spheres then move in the $\pm {\bf\hat{z}}$ direction accordingly. This movement is restricted to displacements $\Delta z_i$, per each sphere, by the restoring force $F_{z,i}=-k \Delta z_i$ [See Fig.~\ref{fig:springmodel}(b)]. The configuration of the deformed objects is no longer $\BR$-even. This might lead to relative velocity, $\Delta V_R=V_{x,1}-V_{x,2}\neq 0$. The displacements in the ${\bf\hat{z}}$ direction will cause the trailing object to be shorter and move faster and the leading object to be longer and move slower. Consequently, the two objects will attract. This heuristic prediction is verified by the results below.

The reversal of the direction of the force makes the two objects switch their roles as leading and trailing [See Fig.~\ref{fig:springmodel}(c)]. As a result, after a transient, the deformation will be reversed as well, $\Delta z_1(-\BF)=\Delta z_2(\BF)$, the velocities will switch sign and exchange, $V_{x,2}(-\BF)=-V_{x,1}(\BF)$, and the relative velocity $\Delta V_R$ will remain the same. Therefore the interaction is $\BF$-even.

The system in Sec.~\ref{sec_wall} and the present one are qualitatively different. While the former consists of wall-bounded rigid spheres, the latter contains unbounded flexible objects. Nonetheless, the analyses of the two problems are very similar, since analogous symmetry-breaking effects are at play.

\subsubsection{Results}
\label{sec_resspring}

\begin{figure}[tbh]
  \centering
  \includegraphics[width=0.5\columnwidth]{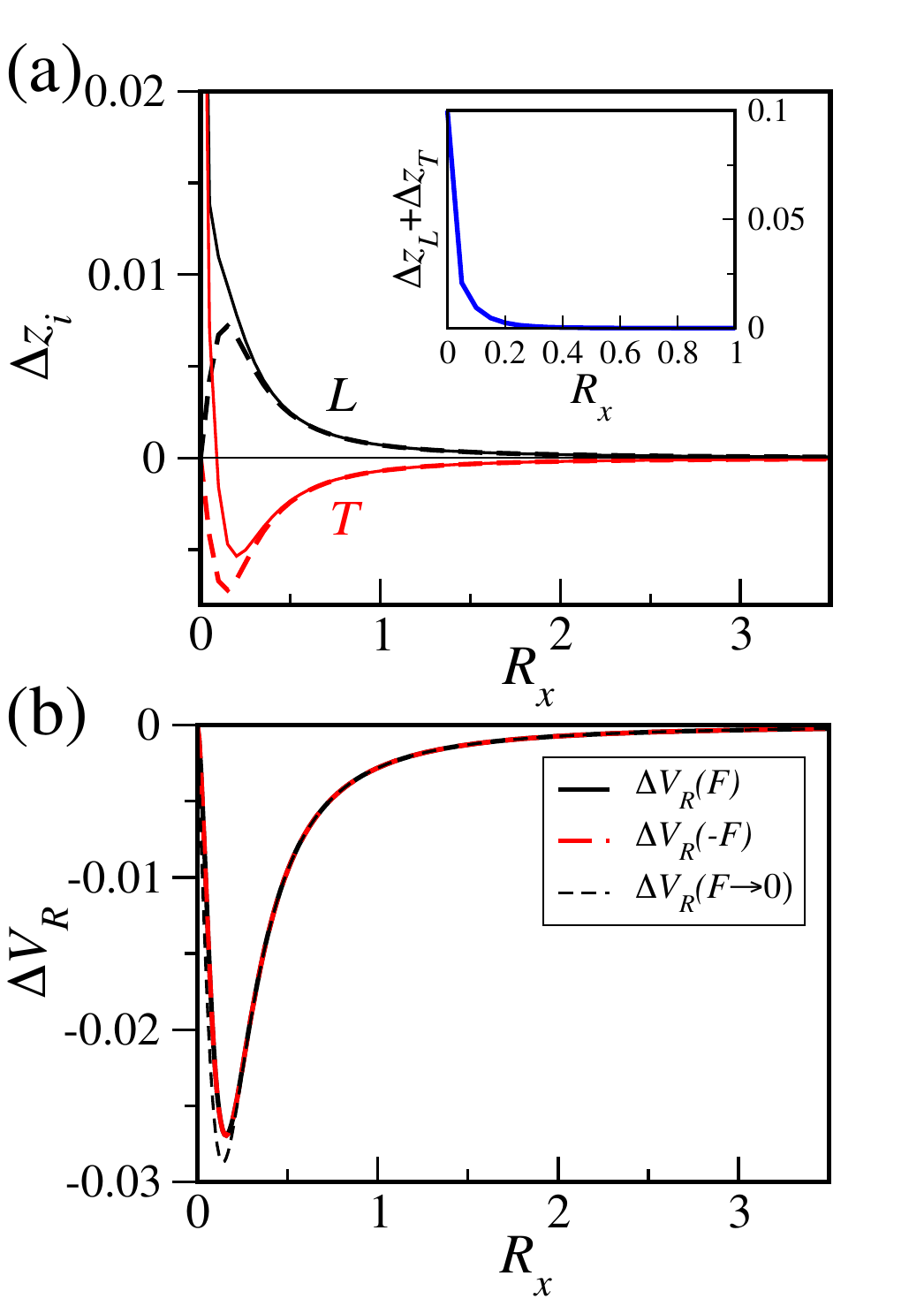}
\caption{(a) The perpendicular displacements of the spheres within the leading ($L$, black) and trailing ($T$, red) objects as a function of the objects' mutual distance. (b) The hydrodynamic interaction (relative velocity) as a function of distance. Dashed lines show the approximation to leading order in weak force. Solid lines show the results to arbitrary order in $F/(kz_0)$, while still assuming small displacements $\Delta z_i$. The sum of the latter  displacements is shown in the inset of panel (a), demonstrating the deviation from antisymmetry. Panel (b) shows also the interaction under force reversal (long-dashed, red line), which exactly overlaps the solid, black line. We use $F=\eta=k\equiv 1$, $a=0.01$ and $z_0=0.1$.}
\label{fig:resspr}
\end{figure}

There are three intrinsic length scales in the problem, $a$, $z_0$ and $F/k$. We calculate the relative velocity to the leading order in particle size, $a \ll z_0, R_x$. The required components of the response tensors can be replaced then in the following manner using the Stokeslet approximation \cite{HappelBook}:
\begin{eqnarray}
&&(\CT_{11})_{xx} \simeq B_{x,1}(z_1),\ \ (\CT_{22})_{xx} \simeq B_{x,2}(z_2),\nonumber \\
&&(\CT_{12})_{xx} \simeq G_{xx}(R_x),\ \ (\CT_{21})_{xx} \simeq G_{xx}(-R_x),\nonumber \\
&&(\CT_{12})_{zx} \simeq G_{zx}(R_x),\ \ (\CT_{21})_{zx} \simeq G_{zx}(-R_x).
\label{sprtensors}
\end{eqnarray}
The self-mobility of each object is a combination of the Stokes mobility of one sphere and its interaction with its partner within the object. Using the Oseen tensor for the interaction we get
\begin{equation}
B_x(z) =\frac{1+(3/8)(a/z)}{6\pi\eta a}.
\label{Bsxspr}
\end{equation}
The hydrodynamic interaction between the two objects can be calculated for any values of $\Delta z_i$, within the Stokeslet approximation (the results of such a calculation are presented graphically in Fig.~\ref{fig:resspr}). For simplicity we assume a strong restoring force such that $\Delta z_i \ll z_0$. To this leading order we can start with two objects of equal length, assuming $z_1=z_2=z_0$. This is equivalent to taking the leading order in $F/(kz_0)$. The hydrodynamic interaction in the ${\bf\hat{x}}$ direction is then
\begin{equation}
G_{xx}(R_x)=G_{xx}(-R_x)=\frac{1}{4\pi\eta}\left(\frac{1}{|R_x|}+\frac{R_x^2}{\rho^3}+\frac{1}{\rho}\right),
\label{Gxxspr}
\end{equation}
where $\rho=\sqrt{R_x^2+4z_0^2}$ is the distance between the diagonal spheres, that is, between the upper (lower) sphere of one object and the lower (upper) sphere of the other object. The velocity of each object in the ${\bf\hat{x}}$ direction, generated by the forces $F$ on itself and on the other object, is
\begin{eqnarray}
  &&V_{x,1}=[B_{x,1}+G_{xx}(R_x)]F,\nonumber \\
  &&V_{x,2}=[B_{x,2}+G_{xx}(-R_x)]F.
  \label{Vxspr}
  \end{eqnarray}
Since $G_{xx}$ is even in $R_x$ [Eq.~(\ref{Gxxspr})], as long as $B_{x,1}= B_{x,2}$, according to Eq.~(\ref{Vxspr}), there is no interaction between the objects in the ${\bf\hat{x}}$ direction \footnote{Similarly to the previous example, $G_{xx}$ is symmetric under the inversion of particle positions also when $z_1 \neq z_2$.}.

In the ${\bf\hat{z}}$ direction, by symmetry, one object does not move the center of mass of the other, $G_{zx}(R_x)=G_{zx}(-R_x)=0$. Yet, each object causes relative motion of the spheres within the other. We denote the upper sphere as $u$ and the lower sphere as $l$. If the spheres were free to move in the ${\bf\hat{z}}$ direction, still assuming objects of equal length $2z_0$, the velocities of the four spheres would be
\begin{eqnarray}
  && V^u_{z,1}=-V^l_{z,1}=G'_{zx}(R_x)F,\nonumber\\
  && V^u_{z,2}=-V^l_{z,2}=G'_{zx}(-R_x)F,
  \label{relVspr}
\end{eqnarray}
where $G'_{zx}$ is the interaction between the diagonal spheres. This interaction, given by
\begin{equation}
G'_{zx}(R_x)=-G'_{zx}(-R_x)=\frac{1}{4\pi\eta}\frac{z_0 R_x}{\rho^3},
\label{Gzxspr}
\end{equation}
is $R_x$-odd, leading to opposite motions between the two upper spheres and the two lower spheres, $V^u_{z,1}=-V^u_{z,2}=-V^l_{z,1}=V^l_{z,2}$ [see Fig.~\ref{fig:springmodel}(b)]. Thus the interaction gives rise to relative velocity of the spheres within each object, $\Delta V_{z,i}=V^u_{z,i}-V^l_{z,i}$, $\Delta V_{z,1}=-\Delta V_{z,2}$. Hence, the leading object becomes longer and the trailing one becomes shorter, as we anticipated in the previous section. This symmetry breaking leads to relative velocity between the objects along the ${\bf\hat{x}}$ direction, $\Delta V_R \neq 0$. The springs' restoring force $F_z$ limits the deformations, and the objects reach a length of $2(z_0+\Delta z_i)$. The steady deformation is determined by the balance between the elastic and hydrodynamic forces on the spheres,
\begin{equation}
  6 \pi\eta a V^u_{z,i}=k2\Delta z_i,
  \label{Vzspr}
\end{equation}
and the same for the lower sphere, by symmetry. Using Eqs.~(\ref{relVspr})--(\ref{Vzspr}), we obtain
\begin{equation}
  \Delta z_L=\frac{3az_0F}{4k}\frac{R_x}{\rho^3},\ \ \Delta z_T=-\Delta z_L.
  \label{dzspr}
\end{equation}
The leading spring gets longer by $2\Delta z_L$, and the trailing gets shorter by $2|\Delta z_T|$. Figure \ref{fig:resspr}(a) (dashed lines) shows the two deformations, $\Delta z_i$, according to Eq.~(\ref{dzspr}). To the leading approximation used in this calculation the  deformations are exactly antisymmetric.

We calculate the velocities in the ${\bf\hat{x}}$ direction [Eq.~(\ref{Vxspr})] by substituting $z=z_0+\Delta z_i$ from Eq.~(\ref{dzspr}) in the self-mobility of Eq.~(\ref{Bsxspr}). To leading order in $\Delta z_i$ this leads to
\begin{equation}
  \Delta V_R=-\frac{3 a F^2}{32 \pi \eta k}\frac{|R_x|}{z_0 \rho^3}.
  \label{dVspr}
\end{equation}
The particles attract as expected. The attraction is nonmonotonous in $R_x$ as seen in Fig.~\ref{fig:resspr}(b) (black dashed line). At sufficiently large $R_x$ the interaction decays as $1/R_x^2$. On the other hand, when $R_x$ decreases relative to $z_0$, the interaction between the diagonal spheres (which is the origin of the attraction between the objects) becomes increasingly weaker relative to the one between the same-side (upper/lower) spheres. As a result the inter-object attraction weakens. The attraction reaches its maximum strength at $R_x=\sqrt{2}z_0$ [black dashed minimum in Fig.~\ref{fig:resspr}(b)]. From the dashed lines in Fig.~\ref{fig:resspr}(a) it would seem that as the two objects get closer, they both eventually return to their relaxed length, $2z_0$. A more accurate calculation (see below) will show otherwise (solid lines). Recall that the whole discussion assumes $a \ll R_x$ which is in line with the parameters of the figure.

Since the objects are made of four small separate spheres (taken as points in the Stokeslet approximation) we can treat them individually taking into consideration higher orders of $F/(kz_0)$. We calculate the hydrodynamic forces on the spheres while assuming nonzero deformations, $\Delta z_i \neq 0$. The force balance of Eq.~(\ref{Vzspr}) then gives two self-consistent equations for $\Delta z_L$ and $\Delta z_T$, which we solve to leading order in $\Delta z_i$. The solid lines in Fig.~\ref{fig:resspr} show the results of this more accurate calculation. The higher orders of $F$ break the antisymmetry of $\Delta z_i$ -- the deformations of the leading and trailing objects differ in magnitude [panel (a) inset]. Below a certain distance $R_x$ the trailing object becomes stretched instead of shrunk. From this distance on, the leading object stretches as well \footnote{The deformation of the leading particle is not necessarily monotonous in $R_x$ as shown in Fig.~\ref{fig:resspr} and depends on the parameters.}.  Upon (nonphysical) contact, $R_x\rightarrow 0$, by symmetry, both objects must have the same length (there is no distinction between leading and trailing). Surprisingly, this length of the `unified' object is not the relaxed one, but stretched. 

Despite all this complexity caused by the higher orders of $F$, the  interaction remains exactly even in the force [panel (b) long-dashed, red line]. This symmetry in $F$ is in line with the expected behavior under force reversal as discussed in the beginning of this example. 

\subsection{Detailed example 3: two spheres driven along a curve}
\label{sec_curve}

\subsubsection{Model}
\label{sec_modelcurve}

\begin{figure}[tbh]
  \centering
  \includegraphics[width=0.4\columnwidth]{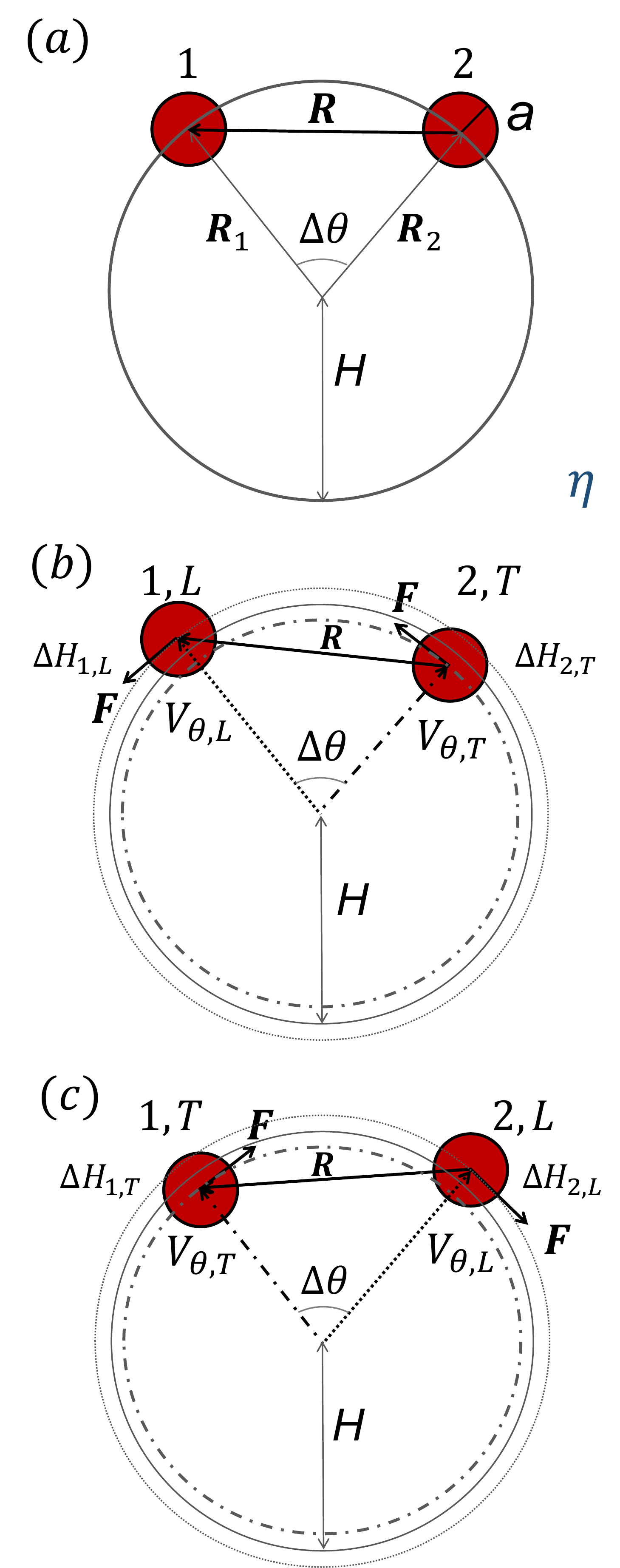}
\caption{Two spherical particles driven along a curved path. (a) In the absence of force the particles are held by an external potential on the same curve of radius $H$. (b) Under the force the particles move in the azimuthal direction and are displaced in the radial direction. The leading particle ($L$) is displaced outwards, and the trailing one ($T$) --- inwards. A restoring harmonic force in the radial direction limits these displacements. (c) Upon the reversal of the force and reaching the new steady state the particles switch roles.}
\label{fig:modelring}
\end{figure}

Taking inspiration from experiments with optical vortices
\cite{Sokolov2011,Sassa2012} and following the same procedure as in the previous examples, we consider a pair of identical spherical particles $1$ and $2$ of radius $a$ positioned on a curve. The curve is taken as a circular arc of radius $\rad$, as shown in Fig.~\ref{fig:modelring}(a). The two particles are separated by the vector $\BR=\BR_1-\BR_2$. We use cylindrical coordinates for the
particle locations, such that $\BR_i=(R_i,\theta_i,z_i=0)$, where $i=1,2$. We denote the angular distance as
$\dtheta=\theta_1-\theta_2$. The particles are driven by a steady azimuthal force $\BF=F\boldsymbol{\hat{\theta}}$ [see
  Fig.~\ref{fig:modelring}(b)]. In response to the force the two particles develop velocities $\BV_i$ with azimuthal components $V_{\theta,i}$. It is known that the particles in such a setup attract hydrodynamically \cite{Sokolov2011}. Here we reexamine this attraction in more detail. We find the relative angular velocity of the particles,
\begin{equation}
\Delta \Omega=\Omega_1-\Omega_2=V_{\theta,1}/R_1-V_{\theta,2}/R_2,
\label{Omega}
\end{equation}
and the symmetries that it obeys. 

When the particles are confined to move strictly on the curve ($R_1=R_2=\rad$), the symmetry under $\BR$-inversion is not broken and therefore there is no relative angular velocity. In other words, to linear order in $\BF$ there is no azimuthal interaction.

The symmetry of this system is broken by the hydrodynamic interaction in the radial direction ${\bf\hat{r}}$, since the azimuthal driving leads to radial motion \cite{Sokolov2011}. This motion in the ${\bf\hat{r}}$ direction is resisted by a restoring force $F_r$, so that at steady state the particles are displaced from their initial radius $\rad$ to $R_i=\rad+\Delta \rad_i$. We assume again a spring-like restoring force, $F_{r,i}=-k \Delta \rad_i$. After the tilt the configuration is no longer $\BR$-even, which allows relative angular velocity, $\Delta \Omega \neq 0$. Specifically, in the tilted configuration the leading particle moves on a curve with a larger radius than that of the trailing one. As a result, the leading particle has a smaller angular velocity and the trailing particle catches up with it. In this way the particles attract. This mechanism was observed in experiments and in Stokesian Dynamics simulations \cite{Sokolov2011}.

Beyond the specific mechanism leading to relative angular velocity, as in the previous two examples, reversal of the force direction makes the two particles switch their roles as leading and trailing [see Fig.~\ref{fig:modelring}(b) and (c)]. In the new steady state, the tilt is reversed, $\Delta \rad_1(-\BF)=\Delta \rad_2(\BF)$, the angular velocities switch sign and exchange, $\Omega_2(-\BF)=-\Omega_1(\BF)$, and the relative angular velocity $\Delta \Omega$ remains the same. Thus the interaction is $\BF$-even. 

Once again, the system described here is physically different from the two previous examples, yet, the qualitative picture is similar. We are going to analyze the present system to higher accuracy, taking into account high orders in $F$ as well as the particles' finite size (going beyond the Stokeslet approximation). In this way we will be able to verify the general symmetry principles for a much more elaborate case which can only be treated numerically.

\subsubsection{Results}
\label{sec_resring}

This system is governed by two dimensionless parameters, $a/H$ and $F/(kH)$. Unlike the previous examples, here we do not assume that $a/H$ is much smaller than unity. We replace the simplest Stokeslet approximation for the interaction with the Rotne-Prager-Yamakawa (RPY) tensor \cite{RotnePrager1969,Yamakawa1970}. The response tensors follow from the RPY tensor according to:
\begin{eqnarray}
&&(\CT_{11})_{\alpha \beta}=(\CT_{22})_{\alpha \beta}=B \delta_{\alpha \beta},\nonumber \\
&&(\CT_{12})_{\alpha \beta}=G_{\alpha \beta}(\BR),\nonumber \\
&&(\CT_{21})_{\alpha \beta}=G_{\alpha \beta}(-\BR),
\label{ringtensors}
\end{eqnarray}
where $B=(6 \pi \eta a)^{-1}$ is the self-mobility, and
\begin{equation}
G_{\alpha \beta}(\BR) =\frac{1}{8 \pi \eta R} \left(\delta_{\alpha \beta} + \frac{R_{\alpha} R_{\beta}}{R^2} \right)+
\frac{1}{12 \pi \eta R} \frac{a^2}{R^2} \left(\delta_{\alpha \beta} - 3 \frac{R_{\alpha} R_{\beta}}{R^2} \right).
\label{RPY}
\end{equation}
We use indices $\alpha,\beta$ to denote the Cartesian coordinates $x,y,z$. Note that the RPY tensor is $\BR$-even.

\begin{figure}[tbh]
  \centering
  \includegraphics[width=0.5\columnwidth]{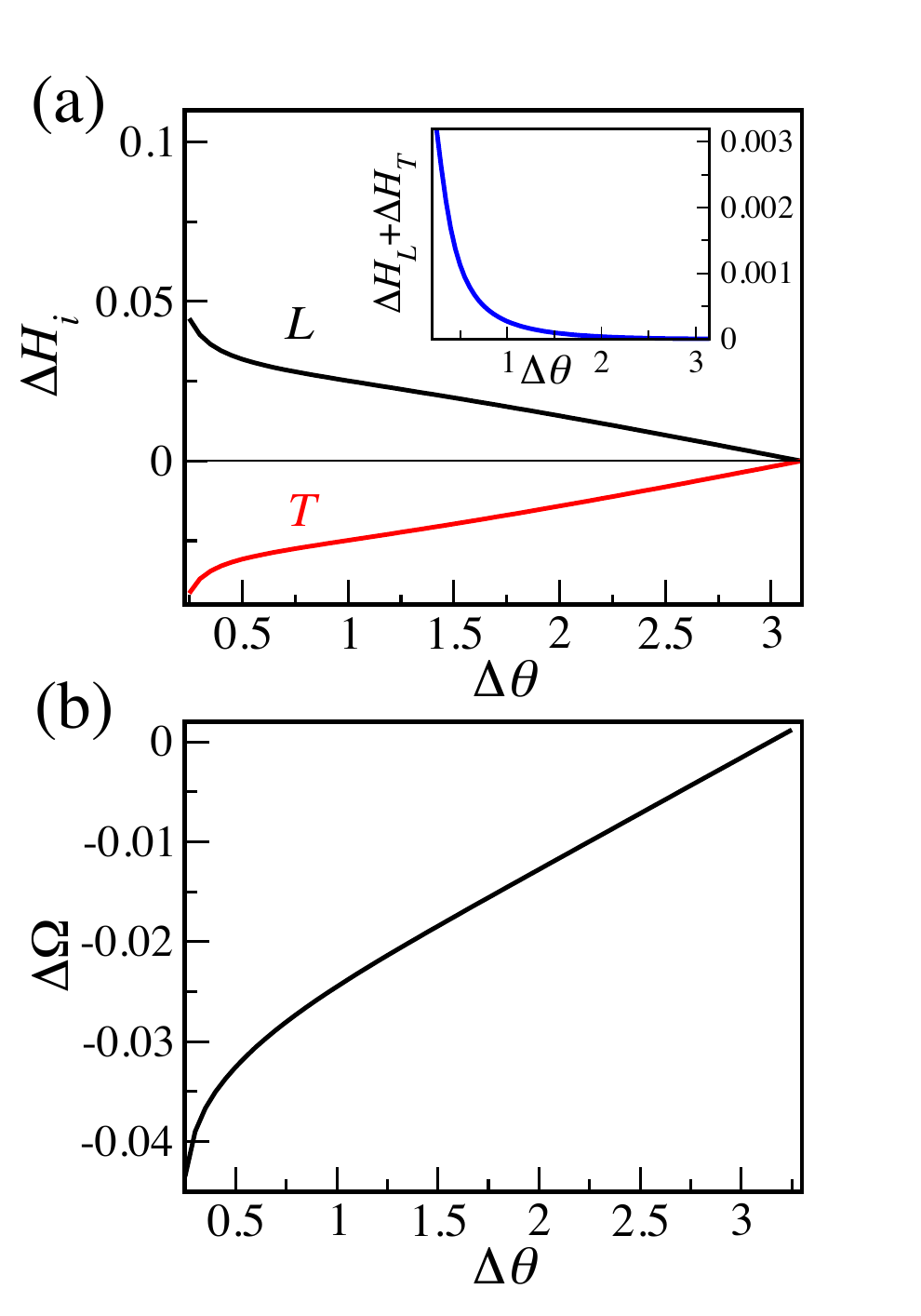}
\caption{Interaction as a function of angular distance. (a) Radial displacements (in $\mu m$) of the leading ($L$, black) and trailing ($T$, red) particles. The inset shows the sum of the two displacements, demonstrating their deviation from antisymmetry. (b) Relative angular velocity (in $rad/s$) shows attraction between the particles. The chosen parameters match those used in the experiment of Ref.~\cite{Sokolov2011}: $a=0.74 \mu m$, $H=6.25 \mu m$, $F=0.25 pN$, $k=1.2 pN/\mu m$ and $\eta=10^{-3}Pa\cdot s$.}
\label{fig:ringOmega}
\end{figure}

In order to calculate the angular velocities we need to find the tangential velocities $V_{\theta,i}$ and the radial positions of the particles $R_i=H+\Delta H_i$. Without any further approximations (beyond RPY), we begin with the already tilted configuration and proceed as follows:

\begin{enumerate}

\item
Since the responses in Eqs.~(\ref{ringtensors}) are given in Cartesian coordinates, we first transform all vectors from cylindrical to Cartesian representation,
\begin{eqnarray*}
&&\BR_i=(R_i \cos \theta_i,R_i \sin \theta_i,0)=(H+\Delta H_i)(\cos \theta_i,\sin \theta_i,0),\\
&&\BF_{\theta,i}=F(-\sin \theta_i,\cos \theta_i,0),\\
&&\BF_{r,i}=F_{r,i}(\cos \theta_i,\sin \theta_i,0).
\end{eqnarray*}
Note that $\Delta H_i$ and $F_{r,i}$ are not known yet.

\item
Subsequently we construct the expressions for the velocities $\BV_i=(V_{x,i},V_{y,i})$, according to Eqs.~(\ref{ringtensors}),
\begin{equation}
V_{\alpha,i}=B \BF_{\alpha,i}+G_{\alpha \beta}(\BR) (F_{\theta,\beta,j}+F_{r,\beta,j}),
\label{ringeq2}
\end{equation}
which depend on the unknown $\Delta H_i$ (the dependence is hidden in $\BR$) and $F_{r,i}$.

\item
Now we project these velocities onto the radial direction,
\begin{equation}
V_{r,i}=\BV_i \cdot (\cos \theta_i,\sin \theta_i,0).
\label{rproj}
\end{equation}

\item
We relate the radial components of the velocities and forces,
\begin{equation}
F_{r,i}=B^{-1} V_{r,i}.
\label{FrV}
\end{equation}

\item
We also relate the radial forces and radial displacements,
\begin{equation}
F_{r,i}=-k \Delta H_i.
\label{FrH}
\end{equation}

\item
  This leads to two self-consistent equations for the two unknowns,
  $\Delta H_1$ and $\Delta H_2$, which we solve numerically. The
  results do not depend on the individual angular positions $\theta_1$
  and $\theta_2$, but rather on their difference $\Delta \theta$, as
  expected from the translation symmetry along the curve. The radial
  tilts increase monotonously with decreasing angular distance [see
    Fig.~\ref{fig:ringOmega}(a)]. In Sec.~\ref{sec_wall}, for example, the obtained
  tilts were antisymmetric, $\Delta h_1=-\Delta h_2$ [see
    Eq.~(\ref{h})], whereas in the present higher-order calculation
  this is true only at large angular distances [see
    Fig.~\ref{fig:ringOmega}(a) inset]. For sufficiently small $\Delta
  \theta$, roughly when the gap between the particles is of the order
  of their diameter, the tilts start growing more sharply, and the
  radial deviation of the leading particle becomes significantly
  larger than that of the trailing one.

\item
We project the velocities onto the azimuthal direction,
\begin{equation}
V_{\theta,i}=\BV_i \cdot (-\sin \theta_i,\cos \theta_i,0).
\label{tproj}
\end{equation}
These expressions depend on $\Delta H_i$ (now known), $\Delta \theta$, $F$ and the rest of the parameters.

\item
Finally, we calculate the relative angular velocity according to Eq.~(\ref{Omega}). 

\end{enumerate}

In Fig.~\ref{fig:ringOmega}(b) we show the numerical results for $\Delta \Omega$ as a function of $\Delta \theta$. As anticipated, we obtain attraction between the particles ($\Delta \Omega <0$). Unlike the previous two examples, here the attraction increases monotonously as the particles approach one another.

\begin{figure}[tbh]
  \centering
  \includegraphics[width=0.5\columnwidth]{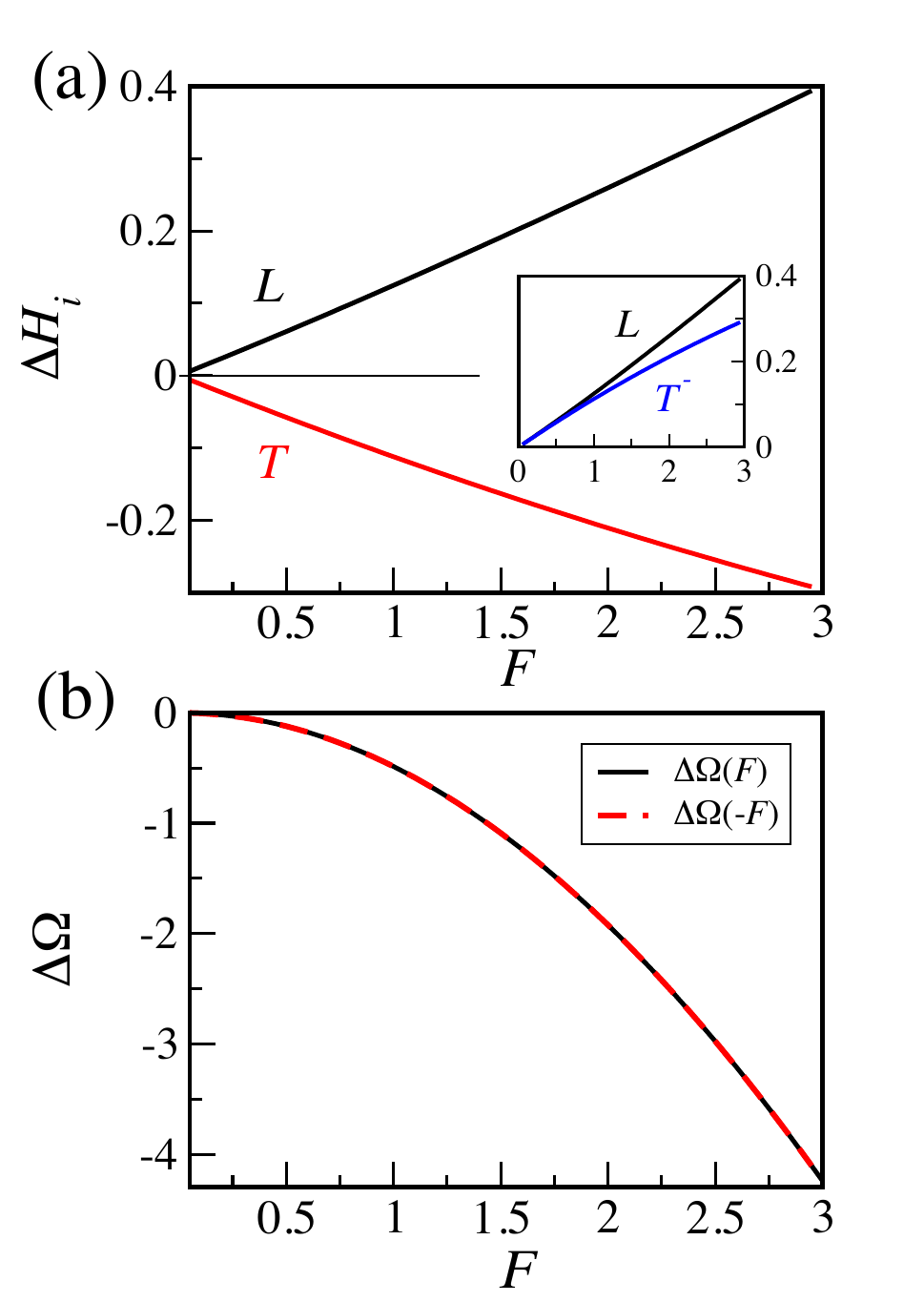}
\caption{Interaction as a function of driving force. (a) Radial displacements (in $\mu m$) of the leading ($L$, black) and trailing ($T$, red) particles. The inset compares the radial displacement of the leading particle ($L$, black) to minus the radial displacement of the trailing particle ($T^-$, blue), demonstrating that they are not antisymmetric. (b) Relative angular velocities (in $rad/s$) versus $F$ (solid black) and $-F$ (dashed red) are indistinguishable. Parameters are the same as in Fig.~\ref{fig:ringOmega}.}
\label{fig:ringF}
\end{figure}

Figure \ref{fig:ringF} shows the dependence of the numerical results for $\Delta H_i$ and $\Delta \Omega$ on the driving force $F$. In Fig.~\ref{fig:ringF}(a) we see that the tilts of the leading and trailing particles are of opposite signs but not exactly antisymmetric for relatively large forces (inset). Nevertheless, Fig.~\ref{fig:ringF}(b) demonstrates that the numerical curves for $\Delta \Omega(F)$ and $\Delta \Omega(-F)$ are indistinguishable. This shows that the interaction is exactly even in $F$, as expected from symmetry above. The obtained response is not quadratic, but contains higher even powers of $F$.

\section{Discussion}
\label{sec_discussion}

In this work we have presented general symmetry arguments concerning the emergence of nonlinear hydrodynamic interactions. We have demonstrated their implications for three detailed examples. The symmetry breaking may lead to either repulsion (first detailed example), or attraction (the other two examples). 

Our symmetry-based analysis and the examples suggest new ways to predict and affect hydrodynamic interactions between colloidal particles and thus direct the dynamics of the particles themselves. Treating a pair of particles with no linear hydrodynamic interaction, e.g., two spheres in an unbounded fluid, we can minimize the nonlinear interaction by using configurations which are close to $\BR$-inversion symmetry. Conversely, we can enhance nonlinear interaction by getting away from this symmetry. The dynamics of the pair, in turn, will affect the collective dynamics.

While the commonly assumed linear dependence on force implies an odd
response to $F$, departure from linear response will generally lead to
$F$-even terms. This will necessarily break time-reversibility, as we
have demonstrated in this work. Consequently, it becomes possible to
apply periodic forces to generate attractive or repulsive interactions
throughout the whole cycle of the periodic drive, making the particles
come closer or further apart without net displacement of the center of
mass. Imagine, for example, a system similar to that described in
Sec.~\ref{sec_wall}, where a layer of particles is held by an external
potential parallel to a wall. When an alternating force is applied
parallel to the wall, nonlinear repulsion will set in and an unbounded
layer should consistently disperse sideways along the axis of the
alternating force. In another scenario the layer is laterally
bounded. Because of the peculiar increase of the repulsive interaction
with increasing distances within a range of $R_x<h$ [see
  Fig.~\ref{fig:reswall}(b)], a concentrated layer (density larger
than $1/h^2$), will undergo an anomalous sharp expansion under an
alternating force. These are qualitative predictions to be verified in
a more concrete treatment of the many-body problem. 

In order to achieve a time-irreversible effect, an additional force, unrelated to the viscous Stokes forces and the drive, must be introduced. In the examples which we have given, these forces came from the elasticity of the deformable objects and boundaries, or the restoring forces due to external potentials.

We would like to underline again the difference between the phenomena discussed here and viscoelasticity. A viscoelastic effect is related to the {\it linear} response to a {\it transient} (finite-frequency) perturbation, whereas here we have dealt with {\it nonlinear} responses to {\it steady} (zero-frequency) forcing. For example, if we take an object constructed of two different spheres connected by a spring and apply a short impulse along the spring, the object will stretch or shrink temporarily and linearly with the force, and then relax back to its initial length through viscous and elastic forces. In the scenario of the present work, i.e., under a steady force, the same object will deform to a steady new length, and will obtain a different self-mobility. Consequently, its ultimate velocity will depend nonlinearly on the force.

There are several extensions to the present work which are worth mentioning. Here we have focused purely on translational interactions, disregarding the rotation of the particles. Rotation is less relevant in the context of this work since linear rotational interaction exists even in the most symmetric configuration of two rigid spheres sedimenting in an unbounded fluid \cite{HappelBook}. Nonlinearities of different origins are another issue to consider. The general considerations presented in Sec.~\ref{sec_gensym} will apply to any nonlinear effect, provided that the two particles reach steady-state velocities which depend only on the force and the configuration. This will not be the case if time-dependent effects, for example, from inertia or viscoelasticity, come into play.

One of the important implications of our results concerns the
collective dynamics of a three-dimensional suspension of many
particles. Pair-repulsions and attractions on the linear level have
been shown to dramatically affect the overall suspension dynamics, the
former leading to suppressed density fluctuations (hyperuniformity)
and the latter to instability (clustering)
\cite{Goldfriend2017,Witten2020}. In the absence of such linear
pair-interactions, applying a similar continuum approach to the
nonlinear interactions treated here will introduce a new nonlinear
coupling between concentration and velocity \cite{Levine1998}. This
calls for a separate study.

\begin{acknowledgments}
This study was supported by the Israel Science Foundation under
Grant Number 986/18.

\end{acknowledgments}

\appendix

\section{Two particles at different distances from a wall}
\label{sec_appA}

In Sec.~\ref{sec_wall} we have shown that there is no linear interaction between particles driven parallel to a wall when their distances from a wall are equal, $h_1=h_2=h$. The interaction arises nonlinearly due to perpendicular tilts of the particles from their initial positions, and the dependence of their mobility on the distance from the wall. It is connected to the fact that the component $G_{xx}$ of Eq.~(\ref{Gxx}) is symmetric under the inversion of their mutual distance $R_x$. Interestingly, the same conclusion holds for a tilted configuration, where $h_1 \neq h_2$. In this case the interaction component becomes \cite{PozrikidisBook}
\begin{equation}
G_{xx}(R_x)=\frac{1}{8 \pi \eta} \left(\frac{\sigma_{-}^2+R_x^2}{\sigma_{-}^3}-\frac{\sigma_{+}^2+R_x^2}{\sigma_{+}^3}-\frac{2h_1 h_2(\sigma_{+}^2-3R_x^2)}{\sigma_{+}^5}\right),
\label{GxxA}
\end{equation}
where $\sigma_{-}=\sqrt{R_x^2+(h_1-h_2)^2}$ is the 3D distance between the particles and $\sigma_{+}=\sqrt{R_x^2+(h_1+h_2)^2}$ is the distance between one of the particles and the ``image'' of the other. This tensor is symmetric under the inversion of particle positions, $R_x \rightarrow -R_x$ and $h_1 \leftrightarrow h_2$.

\end{document}